\documentclass[12pt]{article}
\usepackage{epsfig,amsmath,amsfonts,amssymb,amstext,afterpage,psfrag,
}
\usepackage{slashed}
\usepackage{color}
\usepackage[square,comma,sort&compress,numbers]{natbib}
\usepackage{graphicx}

\allowdisplaybreaks
\usepackage{hyperref}

\newdimen\TW
\usepackage{color}
\definecolor{light-gray}{gray}{0.9}
\definecolor{dark-gray}{gray}{0.7}
\usepackage{tikz}

\long\def\symbolfootnote[#1]#2{\begingroup%
\def\thefootnote{\fnsymbol{footnote}}\footnote[#1]{#2}\endgroup} 

\def\bm#1{\mbox{\boldmath$#1$\unboldmath}}

\def\spose#1{\hbox to 0pt{#1\hss}}
\def\lsim{\mathrel{\spose{\lower 3pt\hbox{$\mathchar"218$}}
 \raise 2.0pt\hbox{$\mathchar"13C$}}}
\def\gsim{\mathrel{\spose{\lower 3pt\hbox{$\mathchar"218$}}
 \raise 2.0pt\hbox{$\mathchar"13E$}}}

\begin{document}

\begin{titlepage}

\begin{flushright}
{\small
LMU-ASC~20/25\\ 
July 2025
}
\end{flushright}

\vspace{0.5cm}
\begin{center}
{\Large\bf \boldmath                                               
  Anomalous Couplings from the Electroweak\\
\vspace*{0.3cm}
  Chiral Lagrangian for Off-Shell Higgs in $gg\to Z_LZ_L$ 
\unboldmath}
\end{center}

\vspace{0.5cm}
\begin{center}
{\sc Gerhard Buchalla and Florian Pandler} 
\end{center}

\vspace*{0.4cm}

\begin{center}
Ludwig-Maximilians-Universit\"at M\"unchen, Fakult\"at f\"ur Physik,\\
Arnold Sommerfeld Center for Theoretical Physics, 
D--80333 M\"unchen, Germany
\end{center}

\vspace{1.5cm}
\begin{abstract}
\vspace{0.2cm}\noindent
We investigate the production of (longitudinal) $Z$-boson pairs
in gluon fusion as a probe of anomalous Higgs couplings.
Of particular interest is the kinematic region of large center-of-mass
energy, where the Higgs-boson is highly off-shell.
We employ the electroweak chiral Lagrangian with a light Higgs,
which is the most natural effective field theory (EFT) for this process.
We demonstrate this by a detailed analysis of the leading and
next-to-leading EFT contributions to the amplitude, 
at leading order in QCD,
emphasizing the role of power counting for a systematic application
of the EFT.
We show that at leading order the new-physics contributions are
described by only two parameters, which depend on three EFT couplings.
Subleading effects can be expected to be small within the range
of validity of the EFT.
Phenomenological implications are briefly discussed.
\end{abstract}

\vfill

\end{titlepage}

\section{Introduction}
\label{sec:intro}

Indirect effects of new physics at the CERN Large Hadron Collider (LHC)
can be pa\-ra\-me\-trized consistently with effective field theory (EFT)
techniques. If the focus is on anomalous Higgs-boson properties, the
most natural framework is the electroweak chiral Lagrangian,
also referred to as nonlinear Higgs-sector EFT, or HEFT
\cite{Feruglio:1992wf,Bagger:1993zf,Koulovassilopoulos:1993pw,Burgess:1999ha,Wang:2006im,Grinstein:2007iv,Contino:2010mh,Contino:2010rs,Buchalla:2012qq,Alonso:2012px,Alonso:2012pz,Pich:2012dv,Buchalla:2013rka,Pich:2013fea,Delgado:2014jda,Buchalla:2015wfa,Buchalla:2015qju,Alonso:2016oah,Pich:2016lew,Pich:2020xzo,Buchalla:2020kdh,Cohen:2020xca,Gomez-Ambrosio:2022qsi,Gomez-Ambrosio:2022why,Delgado:2023ynh}.
By construction, this EFT includes anomalous Higgs couplings
already at leading order. It is organized in terms of loop order
(chiral counting) and defines a systematic, and practical, approximation
scheme for investigating the Higgs sector~\cite{Buchalla:2015wfa}.

In addition to on-shell properties of the Higgs particle in production
and decay, reactions with off-shell Higgs contributions have been
suggested as interesting observables in the literature, in particular
the process of $Z$-boson pair production in gluon fusion, $gg\to ZZ$
\cite{Caola:2013yja,Cacciapaglia:2014rla}.
The topic was analyzed within SM effective field theory (SMEFT)
in~\cite{Azatov:2014jga} and \cite{Ghosh:2025fma}.
Recently, $gg\to ZZ$ has also been addressed in
the context of the electroweak chiral Lagrangian (HEFT)
in~\cite{Anisha:2024xxc}.

In the present paper, we reconsider the EFT treatment of $gg\to ZZ$
within the framework of the electroweak chiral Lagrangian.
We will concentrate on longitudinal $Z$
bosons~\cite{ATLAS:2023zrv,ATLAS:2022oge,Gauld:2017tww,Lyubovitskij:2025oig},
where the Higgs-sector effects can be especially prominent.
We demonstrate how the chiral Lagrangian can be systematically
applied to the partonic process $gg\to Z_L Z_L$, exploiting the
hierarchy among the various anomalous couplings given by the EFT
power counting. This is nontrivial and interesting, since the
chiral (loop) counting of the EFT is intertwined with the
(topological) loop order of the diagrams for $gg\to ZZ$, which
itself is generated at the one-loop level in the Standard Model.
Focusing on the leading-order anomalous effects, a simple and robust
parametrization of Higgs properties is obtained.
Since our emphasis is on the Higgs-boson EFT aspects of
$gg\to Z_L Z_L$, we do not address higher-order QCD corrections,
parton distribution functions or backgrounds. Those are, of course,
important for a full phenomenological analysis.

The present article is the first systematic study of off-shell
Higgs effects in $gg\to ZZ$ within the electroweak chiral Lagrangian
formulation (HEFT). It also serves as an example for the treatment of
similar processes within this EFT framework.

The paper is organized as follows.
In Section~2 we define the electroweak chiral Lagrangian
and present an overview of the contributions to
$gg\to ZZ$ within this framework. We pay particular attention
to the order in the EFT counting, in which the various terms
contribute, distinguishing leading and subleading terms.
Section~3 is devoted to the discussion of the $gg\to Z_LZ_L$
amplitude including the leading anomalous couplings,
and the asymptotic behaviour at large $s$. We also provide
quantitative estimates of next-to-leading order (NLO) corrections
from local operators and from renormalization-group effects.
Phenomenological implications are considered in Section~4.
We conclude in Section~5.
An appendix collects definitions and expressions
for the relevant one-loop functions.

\vspace*{1cm}

\section{\boldmath Effective field theory framework for $gg\to ZZ$}
\label{sec:ewcl}

\subsection{Electroweak chiral Lagrangian}
\label{subsec:ewcl}

The electroweak chiral Lagrangian defines
a weak-scale EFT, extending the Standard Model (SM) by new interactions.
To lowest order, at chiral dimension 2, it is given by
${\cal L}_2={\cal L}_{SM0} + {\cal L}_{Uh,2}$, where
\begin{align}\label{lsm0}
\mathcal{L}_{SM0} = &-\frac{1}{4} G^A_{\mu\nu} G^{A\mu\nu}
-\frac{1}{4} W^a_{\mu\nu}W^{a\mu\nu} 
-\frac{1}{4} B_{\mu\nu}B^{\mu\nu}\nonumber\\
&+\bar q_L i\!\not\!\! Dq_L +\bar\ell_L i\!\not\!\! D\ell_L
 +\bar u_R i\!\not\!\! Du_R +\bar d_R i\!\not\!\! Dd_R
    +\bar e_R i\!\not\!\! De_R 
\end{align}
and
\begin{align}\label{luhlo}
\mathcal{L}_{Uh,2} =& \frac{v^2}{4} \langle D_{\mu} U^{\dag} D^{\mu} U \rangle 
\left(1 + F_U(h)\right) +\frac{1}{2} \partial_{\mu} h \partial^{\mu} h -V(h) 
\nonumber \\ 
&- \Bigl[ \bar q_L \left( {\cal M}_u + \sum_{n=1}^{\infty}  {\cal M}_{u}^{(n)} 
\left( \frac{h}{v} \right)^n  \right) U P_+ q_R   + \bar q_L    
\left( {\cal M}_d + \sum_{n=1}^{\infty} {\cal M}_{d}^{(n)} 
\left(  \frac{h}{v} \right)^n  \right) U P_- q_R  \nonumber \\
& + \bar \ell_L \left( {\cal M}_e + \sum_{n=1}^{\infty} {\cal M}_e^{(n)} 
\left( \frac{h}{v}  \right)^n \right) U P_- \ell_R + \mathrm{h.c.} \Bigr]
\end{align}
with $P_\pm=1/2\pm T_3$ and
\begin{align}\label{fvdef}
F_U(h)=\sum_{n=1}^{\infty} F_n\left(\frac{h}{v}\right)^n\, ,\qquad\quad
V(h)=v^4 \sum_{n=2}^{\infty} V_n\left(\frac{h}{v}\right)^n
\end{align}  
Here $G^A_\mu$, $W^a_\mu$ and $B_\mu$ are the gauge fields
of $SU(3)_C$, $SU(2)_L$ and $U(1)_Y$, respectively, $h$ is the Higgs
field, $v=246\,{\rm GeV}$ the electroweak scale and $\langle\ldots\rangle$
is the $SU(2)$ trace.
The fermions are denoted by $q=(u,d)^T$, $\ell=(\nu,e)^T$, where
the generation indices are suppressed.
The electroweak Goldstone bosons are parametrized
by $U=\exp(2i\varphi^a T^a/v)$ with the $SU(2)$ generators $T^a$
normalized as $\langle T^a T^b\rangle = \delta^{ab}/2$.
Further details on the notation can be found in \cite{Buchalla:2020kdh}.
The Lagrangian ${\cal L}_{SM0}$ represents the unbroken SM.
The scalar sector is described by ${\cal L}_{Uh,2}$, which
includes the leading-order anomalous couplings of the Higgs boson.
The presence of these new-physics effects already in the leading-order
Lagrangian ${\cal L}_2$ is a characteristic feature of the chiral
electroweak EFT.

In the next-order Lagrangian ${\cal L}_4$, of chiral dimension
$d_\chi\equiv 2L+2=4$ (or loop order $L=1$),
several new operators arise, whose list
has been compiled in \cite{Buchalla:2020kdh}. These operators
can be divided into seven classes, denoted by \cite{Buchalla:2020kdh}
$UhD^2$, $UhD^4$, $X^2h$, $XUhD^2$, $\psi^2 UhD$, $\psi^2 UhD^2$, and
$\psi^4 Uh$.
We focus on those operators that can contribute to
the process $gg\to ZZ$ at leading and next-to-leading order
in the chiral counting.
To identify them, we consider all the terms
in the classes above, in particular
\begin{align}\label{quhd2}
  \bm{UhD^2:}\qquad
  Q_{\beta_1} &= v^2 \langle T_3 U^\dagger D_\mu U\rangle^2\, \frac{h}{v}
\end{align}  

\begin{align}\label{quhd4}
  \bm{UhD^4:}\qquad
  Q_{D1} &= \langle D_\mu U^\dagger D^\mu U\rangle^2\, ,\qquad
  Q_{D2} = \langle D_\mu U^\dagger D_\nu U\rangle
           \langle D^\mu U^\dagger D^\nu U\rangle \nonumber\\
  Q_{D7} &= \langle D_\mu U^\dagger D^\mu U\rangle
           \partial_\nu h \partial^\nu h/v^2\, ,\qquad
  Q_{D8} = \langle D_\mu U^\dagger D_\nu U\rangle
           \partial^\mu h \partial^\nu h/v^2 \nonumber\\
  Q_{D11} &= (\partial_\mu h \partial^\mu h)^2/v^4
\end{align}

\begin{align}\label{qx2h}
\bm{X^2h:}\qquad  
Q_{Xh1}&=g^{\prime 2} B_{\mu\nu} B^{\mu\nu}\, \frac{h}{v}\, ,\qquad
Q_{Xh2}=g^2 \langle W_{\mu\nu} W^{\mu\nu}\rangle\, \frac{h}{v}\nonumber\\
Q_{Xh3}&=\frac{g^2_s}{2}\,
                   G^A_{\mu\nu} G^{A\,\mu\nu} \, \frac{h}{v}
\end{align}

\begin{align}\label{qxuhd2}
\bm{XUhD^2:}\qquad  
Q_{XU1}&=g^{\prime}gB_{\mu\nu}\langle W^{\mu\nu}U T_3 U^\dagger\rangle
\, \frac{h}{v}\, ,\qquad 
Q_{XU7}=2ig^{\prime}B_{\mu\nu}
    \langle T_3 D^\mu U^\dagger D^\nu U\rangle \, \frac{h}{v}\nonumber\\
Q_{XU8}&=2ig\langle W_{\mu\nu} D^\mu U D^\nu U^\dagger\rangle \, \frac{h}{v}
\end{align}

\begin{align}\label{qpsiuhd}
\bm{\psi^2 UhD:}\qquad
Q_{\psi V1}&=\bar q_L\gamma^\mu q_L\
  i\langle T_3 U^\dagger D_\mu U\rangle\, ,\qquad
Q_{\psi V2}=\bar q_L\gamma^\mu U T_3 U^\dagger q_L\
  i\langle T_3 U^\dagger D_\mu U\rangle
\nonumber\\
Q_{\psi V4}&=\bar u_R\gamma^\mu u_R\
  i\langle T_3 U^\dagger D_\mu U\rangle\, ,\qquad
Q_{\psi V5}=\bar d_R\gamma^\mu d_R\ i\langle T_3 U^\dagger D_\mu U\rangle
\end{align}

\begin{align}\label{qpsiuhd2}
\bm{\psi^2 UhD^2:}\qquad
Q_{\psi S1}&=\bar{q}_L U P_{+} q_R \langle D_\mu U^\dagger D^\mu U\rangle\, , 
\qquad
Q_{\psi S2}=\bar{q}_L U P_{-} q_R \langle D_\mu U^\dagger D^\mu U\rangle
\end{align}
Finally, there is the class of 4-fermion operators, $\bm{\psi^4 Uh}$,
which we do not list explicitly here.
In writing the operators in (\ref{quhd2}) -- (\ref{qpsiuhd2}) we have assumed
CP conservation, and we have listed the terms with the powers of Higgs
fields $h^n$ relevant for the processes under consideration.
We will further need to consider two operators of the Lagrangian
at NNLO (chiral dimension 6)
\begin{align}\label{qgu}
Q_{GU1} &=\frac{g^2_s}{2}\, G^A_{\mu\nu} G^{A\,\mu\nu} \,
\langle D_\lambda U^\dagger D^\lambda U\rangle\, ,\qquad
Q_{GU2} =\frac{g^2_s}{2}\, G^{A\, \lambda}_{\mu} G^A_{\lambda\nu} \,
\langle D^\mu U^\dagger D^\nu U\rangle
\end{align}
which induce local $ggZZ$ interactions.
Since terms in the electroweak chiral Lagrangian are organized 
by chiral dimensions $d_\chi=2L+2$,
the EFT expansion is constructed by loop orders $L$.
We next discuss at which order in the EFT the operators listed above
enter the amplitude for $gg\to ZZ$.

\subsection{\boldmath EFT applied to $gg\to ZZ$: Overview }
\label{subsec:ampggzz}

\begin{figure}[th!]
\centering
\includegraphics[width=14cm]{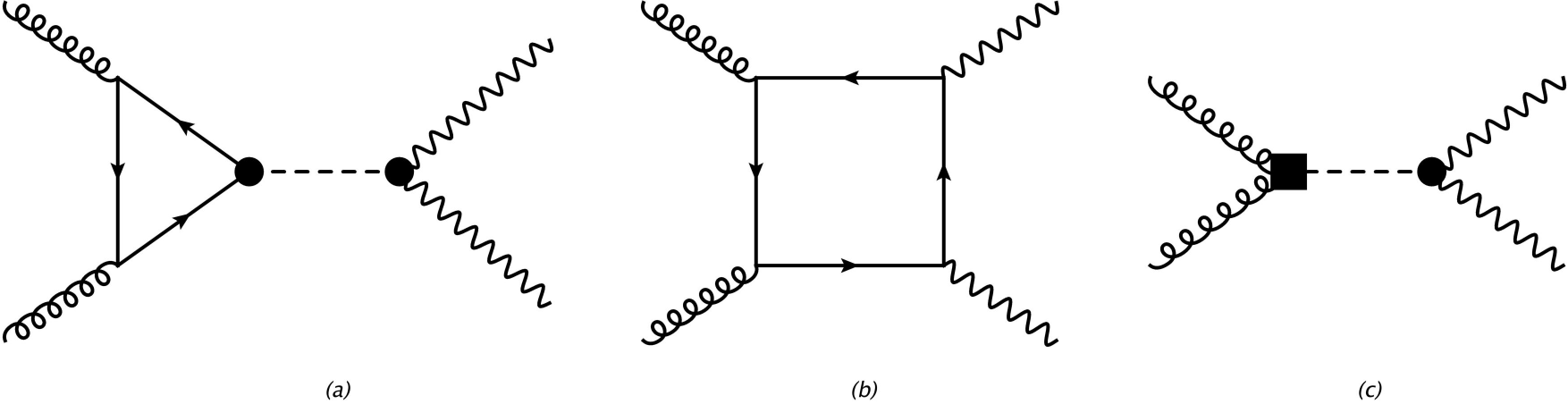}
\caption{\small  Diagrams for $gg\to ZZ$ at leading order in the chiral
  counting.
  Curly, wavy, dashed and full lines refer to gluons, $Z$ bosons,
  Higgs bosons and quarks, respectively.
  Black circles and black squares denote anomalous couplings from
  the LO and NLO Lagrangian, respectively. Additional diagrams with
  permutations of the external legs are not explicitly shown.}
\label{fig:ggzzlo}
\end{figure}

It is important to note that this process is induced at one-loop order,
which therefore defines the leading order in the EFT.
It follows that the diagrams for $gg\to ZZ$ at leading order
are the ones shown in Fig.~\ref{fig:ggzzlo}.
They consist of one-loop topologies $(a)$ and $(b)$, and the tree graph
$(c)$ with a single insertion of a NLO vertex.
Any number of vertices from the leading-order Lagrangian
is allowed in both. 
We remark that anomalous Higgs couplings from the Lagrangian at $d_\chi=2$
in (\ref{luhlo}) enter diagram $(a)$, whereas the gluon-quark and
$Z$-boson-quark interactions in $(b)$ are SM-like at this order.
New physics in $gg\to ZZ$ is then described by three different couplings,
as can be seen from Fig.~\ref{fig:ggzzlo}.

\begin{figure}[th!]
\centering
\vspace*{1cm}  
\includegraphics[width=12cm]{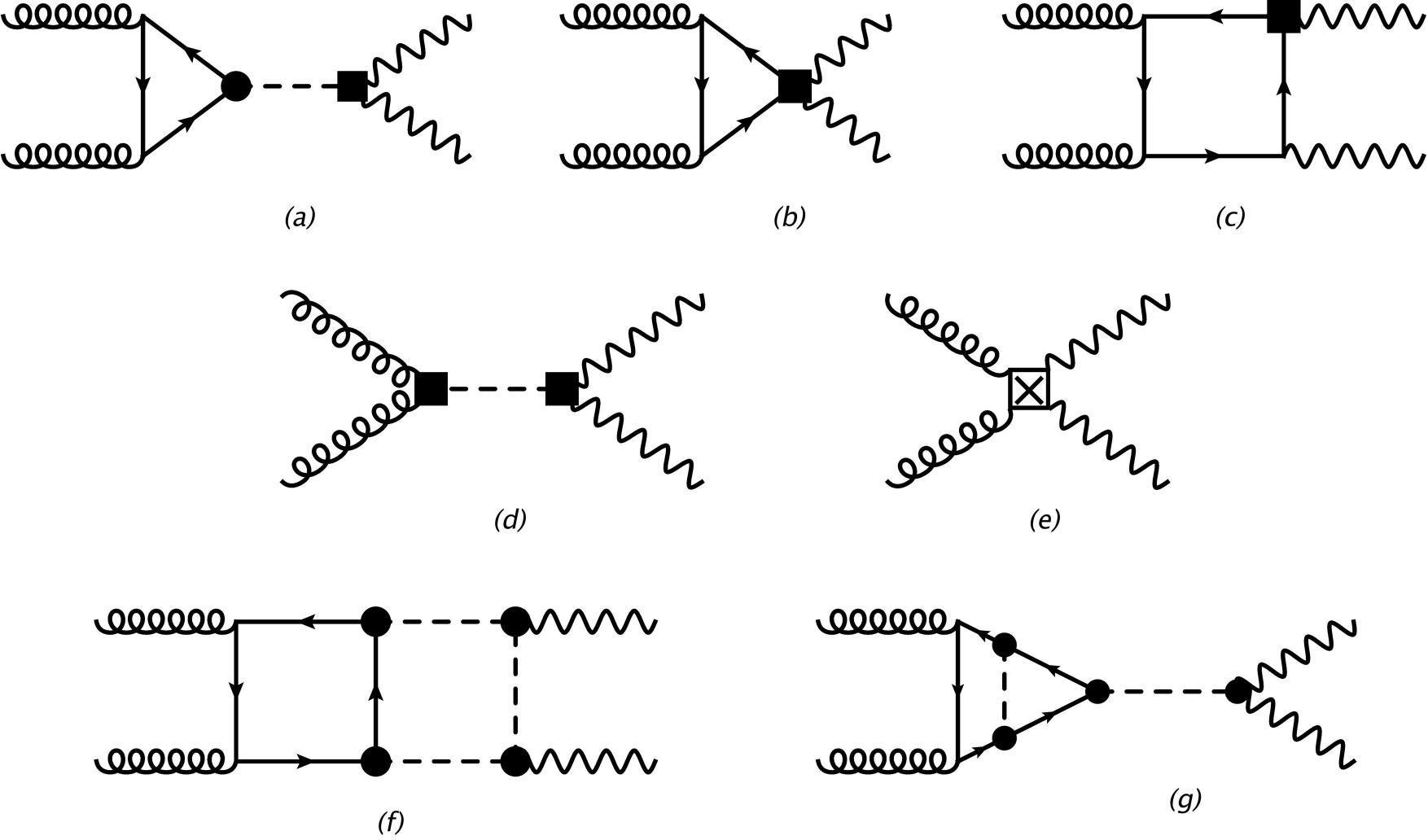}
\caption{\small  Representative diagrams for $gg\to ZZ$ at
  next-to-leading order in the chiral counting.
  Black circles, black squares and crossed squares denote anomalous
  couplings from the LO, NLO and NNLO Lagrangian, respectively.}
\label{fig:ggzznlo}
\end{figure}

We will not attempt to provide a complete calculation of EFT corrections
at next-to-leading order for $gg\to ZZ$.  
However, we present an overview over all contributions that would be
required in such an analysis.
Next-to-leading order in this case means two-loop contributions
in the chiral counting. Representative examples are shown 
in Fig.~\ref{fig:ggzznlo}.
They comprise three classes:
(I) Two-loop topologies with any number of LO vertices, $(f)$ and $(g)$,
(II) one-loop graphs with a single insertion of a NLO vertex
(and any number of LO vertices), $(a)$, $(b)$, $(c)$,
and (III) tree graphs
with either two NLO vertices, $(d)$, or a single NNLO vertex, $(e)$.
Considering the operators in (\ref{quhd2}) -- (\ref{qpsiuhd2}) and
(\ref{qgu}), the following terms may contribute to
the various graphs in Fig.~\ref{fig:ggzznlo}.
\begin{align}\label{qfig2}
\begin{array}{ll}  
 (a):\, Q_{Xh1}, Q_{Xh2}, Q_{XU1}  & (d):\, Q_{Xh3},\, Q_{Xh1}, Q_{Xh2}, Q_{XU1}
  \\
 (b):\, Q_{\psi S1}, Q_{\psi S2}        & (e):\, Q_{GU1}, Q_{GU2}        \\
 (c):\,  Q_{\psi V1}, Q_{\psi V2}, Q_{\psi V4}, Q_{\psi V5}\qquad  &
\end{array}
\end{align}
A NLO effect also arises from $Q_{\beta_1}$ in (\ref{quhd2}).
The corresponding vertex has the same form as the $hZZ$ coupling $c_Z$
at leading order. The coefficient of $Q_{\beta_1}$ can therefore be
absorbed into $c_Z$ as a NLO correction.

\begin{figure}[th!]
\centering
\vspace*{1cm}  
\includegraphics[width=14cm]{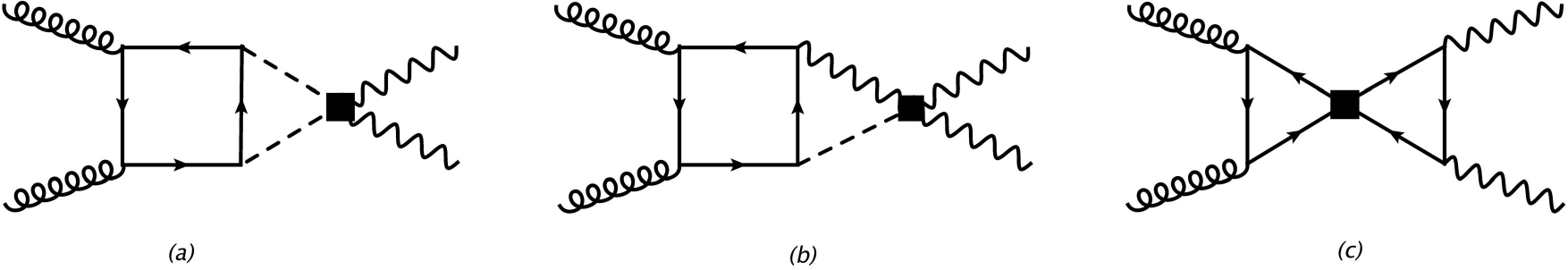}
\caption{\small  Sample diagrams for $gg\to ZZ$ with operators from
  the Lagrangian at chiral dimension 4, which would only
  contribute at next-to-next-to-leading (3-loop) order to this process.
  Black squares denote vertices from the NLO Lagrangian.}
\label{fig:ggzznnlo}
\end{figure}

Finally, there are $d_\chi=4$ operators that contribute to the
$gg\to ZZ$ amplitude only beyond next-to-leading (2-loop) order,
and which should therefore be dropped at this level of approximation.
These are illustrated in Fig.~\ref{fig:ggzznnlo}.
They come from operators in the class $UhD^4$ $(a)$, from
$Q_{XU7}$ and $Q_{XU8}$ of class $XUhD^2$ $(b)$, and from the 4-fermion
interactions of class $\psi^4Uh$ $(c)$.
All graphs in Fig.~\ref{fig:ggzznnlo} are at 3-loop order, with a two-loop
topology and one insertion of a NLO vertex.

QCD corrections at higher order can also be included.
They are governed by the QCD coupling $g_s$ and can therefore
be addressed in perturbation theory, separately from the EFT loop
counting. Since we are mainly concerned with the EFT corrections, we will
not consider higher-order QCD effects
here~\cite{Campbell:2016ivq,Caola:2016trd}.

To summarize the above discussion, we present the couplings for
$gg\to ZZ$ from the chiral EFT in the form of interaction Lagrangians,
which collect the relevant terms at a given order
for this process. The terms from ${\cal L}_2 + {\cal L}_4$
entering the $gg\to ZZ$ amplitude at leading order are
\begin{align}\label{intlo}
{\cal L}_{int,LO} &=  c_V m^2_Z \, \frac{h}{v} Z_\mu Z^\mu
 -m_f c_f\, \frac{h}{v} \bar ff
+\frac{\alpha_s}{8\pi} c_{ggh}\, \frac{h}{v} G^A_{\mu\nu} G^{A\, \mu\nu} 
\end{align}  
At the next-to-leading order we have terms from ${\cal L}_4 + {\cal L}_6$,
\begin{align}
{\cal L}_{int,NLO} &= C_{\beta_1} Q_{\beta_1} + C_{Xh1} Q_{Xh1} + C_{Xh2} Q_{Xh2}
+  C_{XU1} Q_{XU1}\nonumber\\
&+ C_{\psi V1} Q_{\psi V1} + C_{\psi V2} Q_{\psi V2} + C_{\psi V4} Q_{\psi V4}
  + C_{\psi V5} Q_{\psi V5}\nonumber\\
&+ C_{\psi S1}(Q_{\psi S1} + {\rm h.c.}) + C_{\psi S2}(Q_{\psi S2} + {\rm h.c.})
+ C_{GU1} Q_{GU1} + C_{GU2} Q_{GU2} \label{intnlo1}\\
&\supseteq -C_{\beta_1} m^2_Z \, \frac{h}{v}\, Z_\mu Z^\mu  
+\frac{\alpha}{8\pi} C_{ZZh} \frac{h}{v}\, Z_{\mu\nu} Z^{\mu\nu}\nonumber\\
&-\frac{g}{2 c_W} Z_\mu \left( C_{\psi VL}\, \bar t_L\gamma^\mu t_L
   + C_{\psi VR}\, \bar t_R\gamma^\mu t_R\right)
+\frac{g^2}{2c^2_W} C_{\psi S1} Z_\mu Z^\mu\, \bar tt\nonumber\\
&+\frac{g^2_s}{4}\frac{g^2}{c^2_W}\, C_{GU1}\,
G^A_{\mu\nu} G^{A\, \mu\nu}\, Z_\lambda Z^\lambda
+\frac{g^2_s}{4}\frac{g^2}{c^2_W}\, C_{GU2}\,
G^{A\, \lambda}_{\mu} G^A_{\lambda\nu} \, Z^\mu Z^\nu \label{intnlo2}
\end{align}
where the terms with $C_{GU1,2}$ have $d_{\chi}=6$, the others $d_{\chi}=4$,
and
\begin{align}\label{zzhdef}
\frac{\alpha}{8\pi} C_{ZZh} \equiv
g'^2 s^2_W C_{Xh1} +\frac{g^2}{2} c^2_W C_{Xh2} -\frac{g' g}{2} s_W c_W C_{XU1}  
\end{align}
\begin{align}\label{vlrdef}
  C_{\psi VL}\equiv C_{\psi V1}+\frac{1}{2}C_{\psi V2}\, ,
  \qquad C_{\psi VR}\equiv C_{\psi V4}
\end{align}
In (\ref{intnlo2}) we have focused on the top-quark contributions from
the operators with fermions, since those dominate for longitudinal
$Z$-bosons at high energy. 
The coefficients of $Q_{\psi V i}$ modify the $Z$-fermion gauge couplings,
which are constrained by LEP measurements \cite{Han:2004az}.

\subsection{Related EFT studies}

The impact of anomalous Higgs couplings on the process $gg\to ZZ$
within the framework of the electroweak chiral Lagrangian has also
been studied in \cite{Anisha:2024xxc}.
At leading order, \cite{Anisha:2024xxc} includes the effects of $c_t$
and $c_V$, but chooses to omit the local Higgs-gluon coupling $c_{ggh}$.
The latter coefficient can play an important role, in particular for
large center-of-mass energies, as will be discussed below.
At next-to-leading order, \cite{Anisha:2024xxc} considers a set of
14 operators. Of those, assuming weak violation of custodial symmetry,
the 4 operators ${\cal O}_{H8}$, ${\cal O}_{H13}$, ${\cal O}_{d4}$,
${\cal O}_{\Box 0}$, enter only at chiral dimension six and would
contribute only beyond NLO.
The 4 operators
${\cal O}_{HBB}$, ${\cal O}_{HWW}$, ${\cal O}_{H0}$, ${\cal O}_{H1}$,
are in one-to-one correspondence with the terms
$Q_{Xh1}$, $Q_{Xh2}$, $Q_{\beta 1}$, $Q_{XU1}$ in our basis, respectively.
Finally, the 6 operators
${\cal O}_{\Box VV}$, ${\cal O}_{H11}$, ${\cal O}_{d1}$, ${\cal O}_{d2}$,
${\cal O}_{d3}$, ${\cal O}_{\Box\Box}$ can be eliminated in favor of the
other operators in our basis, as we have explicitly checked.
They are therefore redundant. In particular, 
${\cal O}_{\Box VV}$ and ${\cal O}_{\Box\Box}$ can be related to the local
operator $Q_{\psi S1}\sim \bar tt\, Z_\mu Z^\mu$.
In the end, there are only 3 independent operators
($g^2 Z_{\mu\nu}Z^{\mu\nu} h$, $g^2 y_t\bar tt\, Z_\mu Z^\mu$,
$g^3 Z_\mu \bar t\gamma^\mu \gamma_5 t$)
from the $d_\chi=4$ Lagrangian contributing to $gg\to ZZ$ at NLO,
as will be further elaborated in section \ref{subsec:subld} below.

It should also be recalled that the NLO coefficients 
$a_{\Box VV}$ and $a_{\Box\Box}$ from \cite{Anisha:2024xxc}, entering
the EFT at $d_\chi=4$, have a power-counting size of
$a_{\Box VV}\sim a_{\Box\Box}\sim v^2/M^2$, where $M$ is the EFT cutoff scale,
with a typical size of $M\sim 4\pi f\sim 8\, {\rm TeV}$.
Those coefficients may then be expected to be of the order of $10^{-3}$.
The constraints on $a_{\Box VV}$ and $a_{\Box\Box}$ derived in
\cite{Anisha:2024xxc} from current (and projected) LHC data give bounds
of order unity on their values, indicating insufficient sensitivity
on these next-to-leading order couplings.

It is interesting to compare our analysis with a treatment in SMEFT
at dimension 6. At leading order in SMEFT corrections, the anomalous
couplings $c_t$, $c_V$ and $c_{ggh}$ are also generated. Their relation to the
coefficients of operators in the Warsaw basis~\cite{Grzadkowski:2010es}
can be found for instance in~\cite{Buchalla:2018yce}.
We note that the occurrence of $c_{ggh}$ at the same order in the EFT
as $c_t$, $c_V$ rests on the consideration of loop counting
in SMEFT~\cite{Buchalla:2022vjp}.
Concerning $c_t$, $c_V$ and $c_{ggh}$, the pattern of anomalous couplings
is similar in HEFT and SMEFT. However, in SMEFT anomalous $\bar ttZ$
couplings enter in addition at the same order as the latter couplings, in
distinction to HEFT, where nonstandard $\bar ttZ$ effects are subleading.
The process $gg\to ZZ$ has been studied in~\cite{Azatov:2014jga} using
SMEFT and concentrating on the effects from $c_t$ and $c_{ggh}$.
An important feature in SMEFT (different from HEFT) is that deviations
from the SM are parametrically suppressed with the SMEFT cutoff $\Lambda$
by $1/\Lambda^2$ for all anomalous couplings.
An advantage of HEFT is that at leading order the couplings may be
consistently treated as order unity. They can thus be retained without
further expansion in small deviations when squaring the amplitude
and no corresponding ambiguity arises. 

\section{\boldmath Anomalous Higgs couplings in $gg\to Z_LZ_L$}
\label{sec:ggzz}

\subsection{General structure}

In the high-energy limit, we may use Goldstone-boson equivalence
for the $Z$ bosons.
The amplitude for the process $g(k_1,\epsilon_1)\, g(k_2,\epsilon_2) \, \to
\varphi^{0}(p_{1}) \, \varphi^{0}(p_{2})$ can be decomposed as 
\begin{align}
\mathcal{M}^{AB} \, &= \, \delta^{AB} \,
 \mathcal{M}^{\mu \nu}\, \epsilon_{1\mu} \, \epsilon_{2\nu} \,  \label{mab}\\
 \mathcal{M}^{\mu\nu} \, &= \, \frac{\alpha_s}{\pi} \frac{m_t^2}{v^2}
 \left( A_{1}(s,t,u) \, T_{1}^{\mu\nu} +A_{2}(s,t,u) \, T_{2}^{\mu\nu}\right)
\label{ma1a2}
\end{align}
Here $k_1$, $k_2$ are the gluon momenta and $p_1$, $p_2$ the
Goldstone momenta, where
\begin{align}
    k_1^2=k_2^2=p_1^2=p_2^2=0
\end{align}
and we introduced the usual Mandelstam variables 
\begin{align}
    s=(k_1+k_2)^2, \quad t=(k_1-p_1)^2 \quad, u=(k_1-p_2)^2
\end{align}
The amplitudes with longitudinal polarization $\epsilon_L(p)$
of the $Z$ boson at high energy are related to the Goldstone limit through
the replacement $\epsilon^\mu_L(p)\to i\, p^\mu/m_Z$.

$A,B$ are the color indices and $\epsilon_{1\mu}, \epsilon_{2\nu}$ are the
polarization vectors of the gluons.
We define the two linearly independent tensor structures
\begin{align}
T^{\mu\nu}_1 &= g^{\mu\nu}-\frac{2}{s} k^\mu_2 k^\nu_1 \, ,\qquad
T^{\mu\nu}_2 = g^{\mu\nu} + \frac{2s}{tu} p^\mu_1 p^\nu_1 +
 \frac{2}{u} k^\mu_2 p^\nu_1 + \frac{2}{t} p^\mu_1 k^\nu_1
\label{t1t2def}
\end{align}  
which satisfy
\begin{align}
  T_1 \cdot T_2 =0, \quad T_1\cdot T_1 = T_2 \cdot T_2 = 2, \quad
  k_1^{\mu} T^{1,2}_{\mu\nu} = k_2^{\nu} T^{1,2}_{\mu\nu} = 0
\end{align}
The two terms in (\ref{ma1a2}) represent independent helicity configurations.
Amplitude $A_1$ corresponds to gluon helicities of
the same sign ($++$ or $--$), $A_2$ to those of opposite sign ($+-$ or $-+$).
After averaging over spin and color, the squared matrix element reads
\begin{align}
 |\overline{\mathcal{M}}|^2 = \frac{\alpha_s^2}{16 \pi^2} \frac{m_t^4}{v^4}
  \left( |A_1|^2 + |A_2|^2\right)
\end{align}
The differential cross section is then given by
\begin{align}
 \frac{d\sigma}{d\cos\theta} = \frac{|\overline{\mathcal{M}}|^2}{32 \pi s}
\end{align}

\subsection{Form factors at leading order}

The two leading-order form factors are given by 
\begin{align}\label{a1stu}
  A_{1}(s,t,u) = &\frac{1}{s-m_h^2} \left( \left[ 1 -\frac{1}{2} C(s) s
    \left( 1- \frac{4m_t^2}{s}\right)\right]
  \left[s\,(1- c_t c_V) -m_h^2\right]- \frac{c_{ggh}c_V}{4m_t^2} s^2\right)
  \nonumber\\
  &-\frac{1}{2} \left[ 2 + 4 m_t^2 C(s) + s m_t^2
    \left( D(s,t) + D(s,u) + D(t,u)\right)\right]
\end{align}


\begin{align}\label{a2stu}
A_{2}(s,t,u) = &-\frac{1}{4} \frac{1}{tu} \left[ 2s \left(t^2 + u^2\right)
  C(s) + 2 t^3 C(t) + 2 u^3 C(u) - s t^3 D(s,t) - s u^3 D(s,u) \right.
  \nonumber\\
   &\left. + 2 stu \, m_t^2 \, \left( D(s,t) + D(s,u) + D(t,u)\right)\right]
\end{align}

The definitions of the one-loop functions $C$ and $D$ are
collected in appendix \ref{sec:loopf}.
For parts of the calclulation and cross-checks \texttt{Package-X}
\cite{Patel:2015tea,Patel:2016fam} proved useful.
Our results in (\ref{a1stu}), (\ref{a2stu}) are consistent
with~\cite{Glover:1988rg} in the limit $m_Z \to 0$.

\vspace*{0.3cm}

In order to highlight the impact of the anomalous couplings
it is instructive to consider the asymptotic behaviour of
the form factors for large energy $s$.
Keeping the scattering angle fixed, the variables $t$ and $u$
scale with $s$ for $s\to\infty$.
In this limit one finds from (\ref{a1stu}) and (\ref{a2stu})
\begin{align}
  A_1(s,t,u) &=
  -1+\left[-\frac{1}{4} \ln^2\frac{-s}{m^2_t} + 1 \right](1- c_t c_V)
  - c_{ggh} c_V\frac{s+m^2_h}{4 m^2_t} 
  +{\cal O}\left(\frac{1}{s}\right)
\label{a1asy}\\
A_2(s,t,u) &=-\frac{t}{4u} \ln^2\frac{s}{-t} +
i \pi \frac{t}{2u} \ln\frac{s}{-t} +\{ t\leftrightarrow u\}
 +{\cal O}\left(\frac{1}{s}\right)
\label{a2asy}
\end{align}
The asymptotic expression for $A_2$ is of order unity and takes
the special values
\begin{align}
  A_2(s\to\infty)|_{t\to 0} &= i\frac{\pi}{2}\nonumber\\
  A_2(s\to\infty)|_{u=t=-s/2} &=-\frac{1}{2}\ln^2 2 + i\pi \ln 2
  =-0.240 + 2.18 i
\label{a12spc}
\end{align}  
The asymptotic result for $A_1$ in (\ref{a1asy}) reduces to
$A_1\to -1={\rm const.}$ in the SM, where $c_t=c_V=1$, $c_{ggh}=0$,
in agreement with the requirements of unitarity.
A nonzero value of $(1-c_t c_V)$ leads to a term that grows
logarithmically with $s$ \cite{Azatov:2014jga}. By contrast,
a linear increase in $s$ is obtained for the contribution with $c_{ggh}$.
The rise of the amplitude with $s$ is sometimes referred to as
unitarity violation, since it would formally violate unitarity
limits at too high energies. However, unitarity is never actually violated
within the range of validity of the EFT~\cite{Degrande:2012wf}.
Nevertheless, the characteristic $s$-dependence can lead to
relevant experimental signatures in the search for physics beyond
the SM. In particular, the non-standard $s$-dependences in (\ref{a1asy})
have been proposed as a tool to disentangle the Higgs couplings
$c_t$ and $c_{ggh}$, which are difficult to separate in $gg\to h$
with on-shell Higgs~\cite{Azatov:2014jga}.
The impact of this behaviour of the form factors on the partonic cross
section will be further explored in Section~\ref{sec:pheno}.

\subsection{Subleading EFT corrections}
\label{subsec:subld}

Following the discussion in section \ref{subsec:ampggzz},
we now investigate the impact of the local operators
contributing at next-to-leading order to $gg\to Z_L Z_L$.
We take these corrections as representative for the terms at NLO.
We anticipate that these corrections will be sub\-do\-mi\-nant
within the range of validity of the EFT, assuming a generic
size of the coefficients in accordance with the EFT power counting.
In this framework, typical next-to-leading order terms carry a parametric
suppression of
\begin{align}\label{parxim}
\frac{v^2}{M^2} =\frac{v^2}{16\pi^2 f^2}=\frac{\xi}{16\pi^2}
\end{align}  
where $f$ is the scale of the (strongly-coupled) Higgs sector,
$M=4\pi f$ a typical resonance mass acting as the EFT cutoff,
and $\xi\equiv v^2/f^2$. For numerical estimates we will assume
$f\approx 0.7\,{\rm TeV}$ and $M\approx 8\, {\rm TeV}$ as
representative values. The resulting numbers should be understood
as rough order-of-magnitude estimates.
The nonlinear EFT is valid up to energies sufficiently below
the cutoff scale, that is for $s\sim f^2\ll M^2$, in practice
for maximum values of $\sqrt{s}$ around a (few) TeV.
The various operators affect the amplitude in different ways
and we will treat them in turn.

\subsubsection{\boldmath $d_\chi=4$ operator
  $g^2 Z_{\mu\nu} Z^{\mu\nu} h:\, C_{ZZh}$}

For $Z$ bosons with longitudinal polarization, at large $s$, the vertex
$h Z_{\mu\nu} Z^{\mu\nu}$, entering the diagrams in
Fig.~\ref{fig:ggzznlo} $(a)$ and $(d)$, is suppressed relative to the leading
interaction $h Z_\mu Z^\mu$ by a factor of
\begin{align}\label{rzzhv}
R &= -\frac{\alpha}{2\pi}\frac{C_{ZZh}}{c_V}\frac{m^2_Z}{s}
\end{align}  
This implies a correction to the Higgs-exchange contribution
in (\ref{a1stu}), given by the terms with $c_V$, where
\begin{align}\label{cvcorr}
c_V &\to c_V -\frac{\alpha}{2\pi}\frac{m^2_Z}{s} C_{ZZh}
\end{align}  
Since the NLO loop factor has been factored out from our definition
of $C_{ZZh}$, we expect  $C_{ZZh}={\cal O}(1)$. The correction
to $c_V$ in (\ref{cvcorr}) is then negligible, about $4\cdot 10^{-5}$
at $\sqrt{s}=500\, {\rm GeV}$, for longitudinal $Z$.

\subsubsection{\boldmath $d_\chi=4$ operator
   $g^2 y_t Z_\mu Z^\mu \bar tt:\, C_{\psi S1}$}
\label{subsubsec:psis1}

In the Goldstone limit, this interaction amounts to
a term in the Lagrangian of the form
\begin{align}\label{lpsis1}
{\cal L}_{\psi S1} =
 C_{\psi S1}\, \frac{2}{v^2}\partial^\mu\phi^0 \partial_\mu \phi^0\, \bar tt
\end{align}
This vertex contributes through the diagram in Fig.~\ref{fig:ggzznlo} $(b)$
and leads to a correction of the form factor $A_1$ in (\ref{a1stu}) by
\begin{align}\label{delapsis1}
\Delta A^{\psi S1}_1 &=
\left[ 1-\frac{1}{2} C(s) s \left(1-\frac{4 m^2_t}{s}\right)\right](-2s)
\frac{C_{\psi S1}}{m_t}\nonumber\\
&=\left[-\frac{1}{4}\ln^2\frac{-s}{m^2_t} +1\right](-2s)\frac{C_{\psi S1}}{m_t}
+{\cal O}(s^0)
\end{align}
where the last expression is the asymptotic result for large $s$.
To estimate a typical size of the coefficient $C_{\psi S1}$, we
consider a model with a heavy scalar $H$ with Higgs-like couplings
of the form
\begin{align}\label{lhmod}
{\cal L}_H =-\frac{1}{2} M^2 H^2
+\frac{v}{2}\langle D_\mu U^\dagger D^\mu U\rangle H -\frac{m_t}{v}\bar tt\, H
\end{align}  
When integrating out $H$ at tree level, the operator
$Q_{\psi S1}+ {\rm h.c.}$ is generated with a coefficient
\begin{align}\label{cpsis1hmod}
  \frac{C_{\psi S1}}{m_t} = -\frac{1}{2 M^2}
  \sim -\frac{\xi}{16\pi^2} \frac{1}{v^2}
\end{align}
where $\xi=v^2/f^2$ and $M=4 \pi f$, in agreement with the power-counting
expectation.

Another, more specific, scenario would be the extension
of the SM by a heavy scalar singlet in a strong-coupling
regime, as discussed in \cite{Buchalla:2016bse}. In this model, the
EFT coefficient is obtained as $C_{\psi S1}/m_t = -\sin^2\chi/(2 M^2)$.
Here $M$ is the heavy-scalar mass and $\sin\chi$ describes mixing
in the scalar sector, where $\sin\chi={\cal O}(1)$
for the strong-coupling case underlying a nonlinear EFT.
This result is in agreement with the simple estimate in (\ref{cpsis1hmod}).

The correction in (\ref{delapsis1}) shows a linear rise
with $s$ (up to logarithms), similar to the leading-order
contribution with $c_{ggh}$ in (\ref{a1asy}).
It may be incorporated into the leading-order form factor $A_1$
by replacing
\begin{align}\label{cgghpsis1}
c_{ggh} c_V \to c_{ggh} c_V +
\left[\frac{1}{4}\ln^2\frac{-s}{m^2_t} -1\right] \frac{4 m^2_t}{M^2}
\end{align}
where we used (\ref{cpsis1hmod}).
The square bracket in (\ref{cgghpsis1}) is a complex number of order unity,
so that the suppression of the NLO term is given by
$4 m^2_t/M^2\approx 2\cdot 10^{-3}$.

\subsubsection{\boldmath $d_\chi=4$ operators
  $g^3 Z_\mu\bar t\gamma^\mu t_{L,R}:\, C_{\psi VL,R}$}

These operators modify the coupling of $Z$ to top-quarks
and enter the box diagrams as shown in Fig.~\ref{fig:ggzznlo} $(c)$.
In the Goldstone limit, their effect amounts to a correction
factor of
\begin{align}\label{cpsiv}
1+2(C_{\psi VL} - C_{\psi VR})\equiv 1+\delta_V
\end{align}  
for the entire box diagram (the contributions to $A_{1,2}$
in (\ref{a1stu}) and (\ref{a2stu}) without $c_V$).
The correction from $\delta_V$ has an impact on the
terms that grow logarithmically with $s$ in $A_1$, entering the asymptotic
amplitude as
\begin{align}\label{a1dvasy}
  A_1(s,t,u) =
  -\frac{1}{4} \ln^2\frac{-s}{m^2_t} \left[ 1+\delta_V- c_t c_V \right]
  -c_t c_V - c_{ggh} c_V\frac{s+m^2_h}{4 m^2_t} 
  +{\cal O}\left(\frac{1}{s}\right)
\end{align}
$\delta_V$ thus contributes to the logarithmic
growth of $A_1$ in addition to $1-c_t c_V$.
Parametrically, $\delta_V\sim \xi/(16\pi^2)$, which is subleading
to $1-c_t c_V\sim \xi$ and numerically negligible.

The operators in this class may be generated, for example, in a model
with a heavy $Z'$ boson.
We consider a general framework in which the Standard Model (SM)
is extended by a $U(1)'$ symmetry, and the associated $Z'$ boson is a singlet
under the SM gauge group. Different classes of $Z'$ models are distinguished
by the $U(1)'$ charges assigned to the SM fields. The mass of the $Z'$ boson
can arise either via spontaneous breaking of the $U(1)'$ symmetry or through
the Stückelberg mechanism \cite{Ruegg:2003ps}. Discussions of the various
model realizations are given for instance
in~\cite{Carena:2004xs,Dawson:2024ozw,ParticleDataGroup:2024cfk}.
To account for the coupling of the $Z'$ to the Goldstone matrix $U$,
we extend its covariant derivative as follows:
\begin{align}
	D_{\mu} U \rightarrow D_{\mu} U - 2i g_D Q_U \, Z'_{\mu} U T_3
\end{align}
where $g_D$ denotes the $U(1)'$ gauge coupling, and $Q_i = \mathcal{O}(1)$
represent the $U(1)'$ charges of the SM fields.
Following \cite{Dawson:2024ozw}, we consider the general Lagrangian
\begin{align}
  \mathcal{L} = -\frac{1}{4} Z'_{\mu\nu} Z'^{\mu\nu} +
  \frac{M_{Z'}^2}{2} Z'_{\mu} Z'^{\mu} - Z'_{\mu} J^{\mu}
\end{align}
and integrate out the $Z'$ boson at tree level. For simplicity, we neglect
kinetic mixing effects. The current  $J_{\mu}$ is defined as:
\begin{align}
  J_{\mu} = -i g_D Q_U v^2 \langle U^{\dagger} D_{\mu} U T_3 \rangle +
  g_D \left( Q_{t_L} \bar{t}_L \gamma_{\mu} t_L +
  Q_{t_R} \bar{t}_R \gamma_{\mu} t_R \right)
\end{align}
where we retain only the $Z'$ couplings to the Goldstone bosons and the
top quark. Integrating out the $Z'$ yields the effective Lagrangian
\begin{align}
	\mathcal{L}_{\text{eff}} = -\frac{J_{\mu} J^{\mu}}{2 M_{Z'}^2}
\end{align}
which generates the operators $Q_{\psi V 1,4}$ with coefficients
\begin{align}
C_{\psi V 1,4} = \frac{v^2}{M_{Z'}^2} g_D^2 Q_U Q_{t_{L,R}} \sim \frac{\xi}{16\pi^2} 
\end{align}
in agreement with our power counting expectations. Here,
$g_D = \mathcal{O}(1)$ reflects the assumption that SM fermions couple
weakly to the heavy sector.

\subsubsection{\boldmath $d_\chi=6$ operators
  $g^2_s g^2 (G^A)^2\, Z^2:\, C_{GU1,2}$}
\label{subsubsec:gu12}

The $d_\chi=6$ operators lead to a NLO contribution in the form of a purely
local interaction as illustrated in Fig.~\ref{fig:ggzznlo} $(e)$.
The matrix elements read
\begin{align}
  \langle\varphi^0\varphi^0|Q_{GU1}|gg\rangle &=\frac{g^2_s}{2}\delta^{AB}
  \frac{4 s^2}{v^2} T^{\mu\nu}_1 \epsilon_{1\mu} \epsilon_{2\nu}
  \label{qgu1me}\\
  \langle\varphi^0\varphi^0|Q_{GU2}|gg\rangle &=\frac{g^2_s}{2}\delta^{AB}
  \frac{1}{v^2}\left[ -s^2\, T^{\mu\nu}_1 +2ut\, T^{\mu\nu}_2\right]
    \epsilon_{1\mu} \epsilon_{2\nu}
  \label{qgu2me}
\end{align}  
with the notation and definitions from (\ref{mab}) -- (\ref{t1t2def}).
This leads to a form-factor correction of
\begin{equation}\label{delagu}
  \Delta A^{GU}_1=\frac{2\pi^2}{m^2_t}(4 C_{GU1}- C_{GU2})\, s^2\qquad
  \Delta A^{GU}_2=\frac{4\pi^2}{m^2_t} C_{GU2}\, ut
\end{equation}  
We estimate the typical size of the coefficients from a model with a heavy
scalar $H$, similar to (\ref{lhmod}), 
\begin{align}\label{lhmodgu}
{\cal L}_H =-\frac{1}{2} M^2 H^2
+\frac{v}{2}\langle D_\mu U^\dagger D^\mu U\rangle H
+\frac{\alpha_s}{8\pi} c_{ggH} \frac{H}{v}\, G^A_{\mu\nu} G^{A\mu\nu}
\end{align}
which gives
\begin{equation}\label{cgumod}
  C_{GU1}=\frac{c_{ggH}}{32\pi^2 M^2}\sim \frac{1}{16\pi^2 M^2}\, ,\qquad
  C_{GU2}=0
\end{equation}  
This shifts $A_1$ by
\begin{equation}\label{dela1guh}
\Delta A^{GU,H}_1 =\frac{s}{4 m^2_t}\frac{s}{M^2}c_{ggH}
\end{equation}  
which can be phrased as a correction to the $c_{ggh} c_V$ term in $A_1$
\begin{equation}
c_{ggh} c_V\to c_{ggh} c_V - c_{ggH}\frac{s}{M^2}
\end{equation}  
The relative correction is of order $s/M^2\sim f^2/M^2$ and again small.
We remark that the contact interactions in Fig.~\ref{fig:ggzznlo} $(e)$
lead to an intriguing quadratic dependence on $s$ (\ref{delagu}),
stronger than the leading-order $s$-dependence.
Nevertheless, as we have seen, these contributions remain
subleading for large $s$,
as long as we stay within the range of validity of the EFT.
Conversely, for smaller values of $s$, the quadratic dependence
implies a particularly severe suppression.
In the context of SMEFT, a local $ggZZ$ coupling from a dimension-8
operator similar to $Q_{GU1}$ has also been discussed
in~\cite{Azatov:2014jga}, with similar conclusions.

\subsection{\boldmath Toy models for $c_{ggh}$}
\label{subsec:toycggh}

It is instructive to consider how the local coupling $c_{ggh}$ is
generated from the exchange of heavy resonances in some toy scenarios.
In a first simple example we assume the existence of
a heavy vector-like quark $Q$ with Lagrangian
\begin{align}\label{lvquark}
{\cal L}_Q = \bar Q i\!\not\!\! DQ - M_Q \bar QQ - y\, \bar QQ\, h
\end{align}
where $M_Q\approx 4\pi f$ is the resonance mass, and $y\approx 4\pi$
is the (strong) coupling of the Higgs boson to the vector-quark.
Such a vector-like quark generates $c_{ggh}$ through triangle diagrams
at one loop, in analogy to the top-quark.
Retaining the full $s$-dependence, the result can be expressed as
\begin{align}\label{cgghmq}
  c_{ggh} \to c_{ggh}(s) = \frac{y \, v}{M_Q} \, F_Q(s) =
  \frac{2}{3}\frac{y \, v}{M_Q} + {\cal O}(s)
\end{align}
\begin{align}\label{fqtau}
  F_Q(s) = \frac{1}{\tau}
  \left[1+\left(1-\frac{1}{\tau}\right) \arcsin^2 \sqrt{\tau} \right]
  =\frac{2}{3} + \frac{7}{45}\tau + \frac{4}{63}\tau^2 + \mathcal{O}(\tau^3)
\end{align}
where $\tau = s / 4 M_Q^2$.
The local EFT-coupling $c_{ggh}$ in this model is given by the
limit $s=0$ in (\ref{cgghmq}). Parametrically it is of order
$c_{ggh}\sim v/f=\sqrt{\xi}$.
As is well known from the Higgs-gluon coupling in the SM induced by a
top-quark loop, the local approximation, corresponding to the
$\tau=0$ limit, is rather reliable even for sizeable values
of $\tau$ \cite{Djouadi:2005gi}. 
For example, $F_Q(0)=0.6667$ is only changed to $F_Q(1/4)=0.7101$
at an energy as large as $s=M^2_Q$, which is already outside the range
of validity of the EFT. For lower energies, where the EFT holds,
say at $\sqrt{s}=1\, {\rm TeV}$, and taking $M_Q=8\, {\rm TeV}$,
we have $F_Q(\tau)=0.6673$, which is basically unchanged compared
to the $s=0$ limit.
This observation implies that the treatment of
the Higgs-gluon coupling from new physics by a local operator
with coefficient $c_{ggh}$ should be an excellent approximation
throughout the range of validity of the EFT.

As an additional example,
we consider a model with a heavy, colored scalar $S$ in a representation
$R$ of $SU(3)$ coupled to the Higgs singlet $h$
\begin{align}\label{lscalar}
  \mathcal{L}_S =
  D_{\mu}S^{\dagger}D^{\mu}S - M_S^2 S^{\dagger}S - \kappa \, S^{\dagger}S\, h 
\end{align}
where the color indices have been suppressed.
We allow $\kappa \sim 4\pi M_S$ in the strong-coupling case. 
The covariant derivative here is given by
\begin{align}
	D_{\mu}S = (\partial_{\mu} + i g_s T^A_{R} G^A_{\mu}) S
\end{align}
where $T^A_{R}$ are the $SU(3)$ generators in the representation $R$.
The effects of integrating out $S$ at the one-loop level on the
leading order form factor $A_1$ are taken into account by replacing 
\begin{align}\label{cgghms}
	c_{ggh} \to c_{ggh}(s)=\frac{\kappa \, v}{M_S^2} T(R) \, F_S(s)
\end{align}
Here $T(R)$ is the index of the representation $R$ and the loop function is
\begin{align}\label{fstau}
  F_S(s) = \frac{1}{2 \tau} \left( \frac{\arcsin^2 \sqrt{\tau}}{\tau}-1\right)
  = \frac{1}{6} + \frac{4\tau}{45} + \frac{2\tau^2}{35} + \mathcal{O}(\tau^3)
\end{align}
where $\tau = s / 4 M_S^2$.
To get the matching result for the local operator $c_{ggh}$, only the leading
term in the expansion in $\tau$ is relevant. For a scalar octet
($T(\textbf{8})=3$) we get
\begin{align}
	c_{ggh} = \frac{\kappa\,v}{2 M_S^2} \sim \sqrt{\xi} 
\end{align}
which agrees with our power counting expectation.

\subsection{\boldmath Renormalization group running}
\label{subsec:rgrun}

Another class of NLO contributions arises from the renormalization‐group (RG)
running of the leading‐order anomalous couplings.
The one‐loop renormalization of the Higgs–Electroweak Chiral Lagrangian
has been computed in \cite{Buchalla:2020kdh,Buchalla:2017jlu}
(see also \cite{Alonso:2017tdy}).
At this order, the beta function
for \(c_{ggh}\) vanishes, so we only need to consider the RG evolution
of \(c_V\) and \(c_t\). 

The beta function of coefficient $c_i$ is defined by
\begin{align}\label{betaci}
	\beta_{c_i} = 16\pi^2 \frac{d c_i}{d \ln \mu}
\end{align}
At one loop $\beta_{c_V}$ and $\beta_{c_t}$ read
(retaining only the top-quark contribution from the Yukawa sector)
\begin{align}\label{betacv}
  \beta_{c_V} &= \frac{3}{8} \frac{v^2}{m_h^2} c_V (c_V^2-c_{2V})
  (3g^4+2g^2g'^2+g'^4)+\frac{g^2}{12}  c_V
  \left[ 37 (c_{2V}-c_V^2)+17(1-c_{2V})\right] \nonumber \\
  & + \frac{3}{4} g'^2 c_V(1-c_V^2) + \frac{m_h^2}{2v^2}
  \left[ c_V \left( 10(c_V^2-c_{2V})+4(c_{2V}-1)\right)+ 6c_{3V}\right]
  \nonumber \\
  & + 24 \frac{m_t^4}{m_h^2 v^2} c_t (c_{2V}-c_V^2)+
  6 \frac{m_t^2}{v^2} (c_V-1)(c_t^2+1)+ 6 \frac{m_t^2}{v^2} (c_t-1)^2 
\end{align}
and
\begin{align}\label{betact}
  \beta_{c_t} &= \frac{3}{8} \frac{v^2}{m_h^2} c_V [c_t(c_t-c_V)-2c_{2t}]
  (3g^4+2g^2g'^2+g'^4)+\frac{17g^2+9g'^2}{12} c_t (1-c_V^2) \nonumber \\
  & +\frac{m_h^2}{v^2} \left[ c_t \left( 3 \kappa_3 (c_t-c_V) +
    2 c_V^2 +c_{2V} -2 c_{2t}\right)-3c_V+ 6c_{3t}\right]  \nonumber \\
  & + 24 \frac{m_t^4}{m_h^2 v^2} c_t \left[ 2 c_{2t} + c_t(c_V-c_t)\right]+
  6 \frac{m_t^2}{v^2} c_t (c_t^2-1+2c_{2t})
\end{align}
Note that all three couplings, $c_V$, $c_t$ and $c_{ggh}$, are scale
invariant under QCD. This simplifies their interpretation in the
presence of QCD radiative corrections.

We employ the definitions
\begin{align}
  &F_1=2 c_V, \quad F_2 =c_{2V}, \quad  F_3=c_{3V},
 \quad V_2  =\frac{m_h^2}{2v^2}, \quad V_3 =\frac{m_h^2}{2v^2} \kappa_3
  \nonumber\\
  &\mathcal{M}_t = m_t, \quad \mathcal{M}^{(1)}_t = c_t m_t,
  \quad \mathcal{M}^{(2)}_t = c_{2t} m_t, \quad \mathcal{M}^{(3)}_t = c_{3t} m_t
\end{align}
which relate the parameters of the Lagrangian in (\ref{luhlo})
to the phenomenological couplings.
Both $\beta_{c_V}$ and $\beta_{c_t}$ vanish in the SM-limit 
\begin{align}\label{smlim}
c_V=c_{2V}=c_t=\kappa_3=1,\qquad c_{3V}=c_{2t}=c_{3t}=0  
\end{align}  

With numerical values for the parameters, the beta functions
in (\ref{betacv}) and (\ref{betact}) become
\begin{align}\label{bcvnum}
  \beta_{c_V} &= 22.69 c_t(c_{2V}-c^2_V) -5.92 c_t + 3.14 c_V
  + 2.96 c^2_t c_V\nonumber\\
  &- 1.03 c_{2V} c_V + 0.849 c^3_V + 0.773 c_{3V} 
\end{align}  
\begin{align}\label{bctnum}
  \beta_{c_t} &=50.78 c_t c_{2t} + 23.65 c^2_t c_V -19.73 c^3_t
  -2.27 c_t + 0.258 c_t c_{2V} - 1.93 c_V c_{2t}\nonumber\\
  &- 1.15 c_t c^2_V +1.55 c_{3t} +0.773 (c_t(c_t-c_V)\kappa_3 - c_V)
\end{align}  
The large numerical coefficients are dominated by the
terms in (\ref{betacv}) and (\ref{betact}) carrying
a $m^4_t$-dependence, which are formally leading in the limit of
large top-quark mass. 


Solving the RG equation (\ref{betaci}) to linear order in the
beta functions we have 
\begin{align}\label{cvtrun}
  c_{i}(\mu_1) = c_{i}(\mu_2) + \frac{\beta_{c_{i}}}{16 \pi^2}
  \ln \frac{\mu_1}{\mu_2}
\end{align}
We imagine a scenario where the new physics resides at a scale of
$\mu_1=8 \, \text{TeV}$ (the EFT cutoff),
whereas the coefficients $c_V$ and $c_t$
are determined in experiments at a scale of $\mu_2 = 1 \, \text{TeV}$.
Numerically, retaining only the $m^4_t$ terms in the beta functions
(\ref{betacv}), (\ref{betact}), we find for the evolution in (\ref{cvtrun})
\begin{align}
c_V(\mu_1) &\approx c_V(\mu_2)+ 0.30 \, c_t(c_{2V}-c_V^2) \\
c_t(\mu_1)  &\approx c_t(\mu_2)+ 0.30 \,c_t \left[ c_t (c_V-c_t)+2c_{2t}\right]  
\end{align}
These results indicate that RG running effects could have a sizable impact
on the anomalous couplings, when comparing their values at the scale
of LHC measurements with those at the EFT cutoff. 
Experimentally $c_t$ and $c_V$ are close
to one (within 10\%) \cite{deBlas:2018tjm}, but $c_{2t}$ and $c_{2V}$ could
still deviate from their SM values in (\ref{smlim}).

\section{Phenomenological considerations}
\label{sec:pheno}

\begin{figure}[h!]
	\centering
	\includegraphics[width=0.75\textwidth]{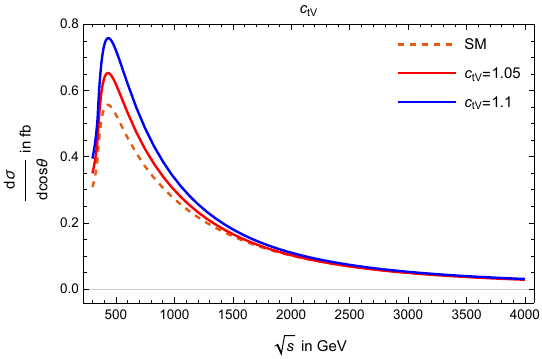}
        \caption{Energy dependence of the scattering cross section at
       $\cos \theta =0$ in units of $fb$. Here only the HEFT coefficient
       $c_{tV}\equiv c_t c_V$ is varied while all other coefficients are zero.}
\label{fig:ctv}
\end{figure}

\begin{figure}[h!]
	\centering
	\includegraphics[width=0.75\textwidth]{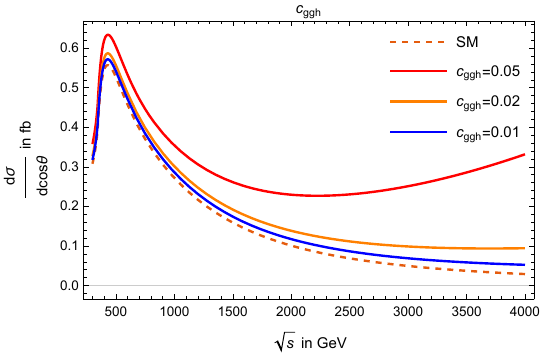}
	\caption{Energy dependence of the scattering cross section at
          $\cos \theta =0$ in units of $fb$. Here only the HEFT coefficient
          $c_{ggh}$ is varied while all other coefficients are zero.}
\label{fig:cggh}
\end{figure}

\begin{figure}[h!]
	\centering
	\includegraphics[width=0.75\textwidth]{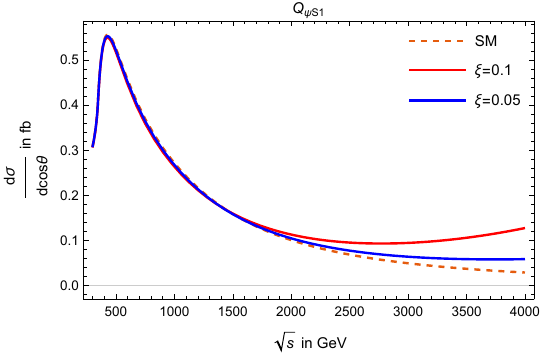}
	\caption{Energy dependence of the scattering cross section at
     $\cos \theta =0$ in units of $fb$. Here only the HEFT coefficient
    $C_{\psi S1}=-\xi m_t/16\pi^2 v^2$ is varied while all other coefficients
      are zero.}
\label{fig:qpsis1}
\end{figure}

\begin{figure}[h!]
	\centering
	\includegraphics[width=0.75\textwidth]{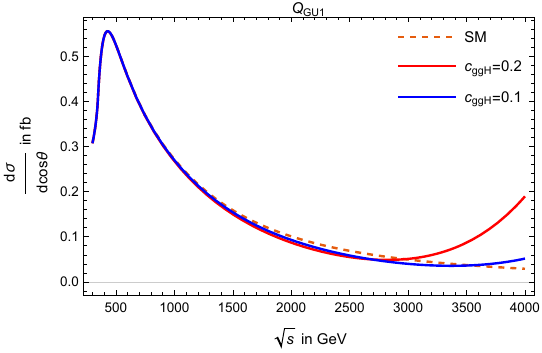}
	\caption{Energy dependence of the scattering cross section at
  $\cos \theta =0$ in units of $fb$. Here only the HEFT coefficients
 $C_{GU1}=c_{ggH}/32\pi^2 M^2$ is varied while all other coefficients are zero.}
\label{fig:qgu1}
\end{figure}

\begin{figure}[h!]
	\centering
	\includegraphics[width=0.75\textwidth]{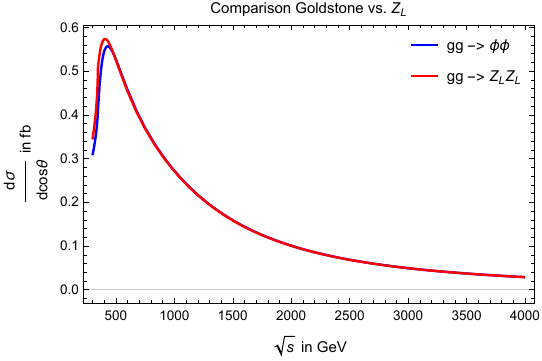}
	\caption{Energy dependence of the scattering cross section at
  $\cos \theta =0$ in units of $fb$. Here we plot the SM scattering cross
  section for the processes $gg \to \varphi^0 \varphi^0$ and $gg \to Z_L Z_L$.
  For large $\sqrt{s}$ we find good agreement between the two processes
  validating our choice to use the Goldstone limit. }
\label{fig:goldstonezl}
\end{figure}

In this section we illustrate the impact
of anomalous couplings in an exploratory analysis.
Concretely, we investigate the partonic differential cross section
\begin{align}
	\frac{d\sigma}{d\cos\theta}\Bigg|_{\theta=\pi/2}
\end{align}
as a function of the partonic center-of-mass energy \(\sqrt{s}\), varying
one anomalous coupling at a time and setting all others to their SM values.

Our primary goal is to elucidate the EFT systematics rather than to perform
a detailed phenomenological study. This would require in particular
the inclusion of important QCD effects ($K$ factors), which is beyond the
scope of the present paper. A systematic treatment of QCD corrections
in the context of the electroweak chiral Lagrangian has been performed
for instance in \cite{Buchalla:2018yce,Buchalla:2022igv,Braun:2025hvr}.

Our input parameters are collected in Table \ref{tab:inputparam}.
\begin{table}[h!]
	\centering
	\begin{tabular}{|c|c|c|c|c|}
	\hline
	$m_t$ & $m_Z$ & $m_h$& $\alpha_s(m_Z)$ & $G_F=1/\sqrt{2}v^2$ \\ \hline
	$173 \, \text{GeV}$   & $91.19 \, \text{GeV}$   & $125 \, \text{GeV}$
        & $0.1179$   & $1.166 \cdot10^{-5} \, \text{GeV}^{-2}$   \\ \hline
	\end{tabular}
	\caption{Input parameter used in the analysis taken from
          \cite{ParticleDataGroup:2024cfk}}
	\label{tab:inputparam}
\end{table}
The numerical evaluation of the loop functions in our analysis was performed
using the \texttt{LoopTools} package \cite{Hahn:1998yk}.\\
While the NLO operators remain largely unconstrained,
beyond power-counting estimates, we can be more specific
about the LO anomalous couplings. Those have been extracted in a
global fit~\cite{deBlas:2018tjm}
\begin{align}
  c_V = 1.01 \pm 0.06, \quad c_t = 1.01^{+0.09}_{-0.10} , \quad
  c_{ggh} = -0.01^{+0.08}_{-0.07} 
\end{align}
where the error bars correspond to the 68 \% probability interval.
Therefore, the combination $c_{tV}=c_tc_V$ can still deviate from unity by
roughly {10\%}.
For the NLO operators we use our power counting expectations,
discussed in sections \ref{subsubsec:psis1} and \ref{subsubsec:gu12},
\begin{align}\label{cpsis1cgu1}
  C_{\psi S1} \sim -\frac{\xi}{16\pi^2} \frac{m_t}{v^2},
  \quad C_{GU1}\sim \frac{c_{ggH}}{32\pi^2 M^2}
\end{align}
Results are displayed in Figs.~\ref{fig:ctv} -- \ref{fig:goldstonezl}. 
We make several remarks:

\begin{itemize}
\item In Fig.~\ref{fig:ctv} we analyze the effect of the LO coupling
  $c_{tV}=c_tc_V$. These couplings impact the cross-section most strongly for
  small $\sqrt{s}\sim500 \text{GeV}$. The relative $\ln^2 s$ growth with
  respect to the SM amplitude only becomes noticeable for center-of-mass
  energies far outside the range of validity of the EFT.
\item The leading new-physics effect at larger $\sqrt{s}$ is due to
  $c_{ggh}$ as can be seen in Fig.~\ref{fig:cggh}. For values of
  $c_{ggh}$ close to central values of the global fit, we remain within the
  range of validity of the EFT. The deviations from the SM increase for
  large energy, proportional to $s$ at the amplitude level.
\item The corrections to the cross-section coming from $Q_{\psi S1}$ are
  illustrated in Fig.~\ref{fig:qpsis1}. As discussed in
  section~\ref{subsubsec:psis1}, they are qualitatively similar to the
  leading effect from $c_{ggh}$, increasing linearly in $s$ at the
  amplitude level, up to a factor containing $\ln^2 s$.
  However, the correction from $Q_{\psi S1}$ is a NLO term and thus
  parametrically suppressed by $\xi/16\pi^2$.
  Numerically, the suppression is somewhat weakened by the $\ln^2 s$
  term in (\ref{cgghpsis1}) and it also depends on the size
  of the coefficient $C_{\psi S1}$. In Fig.~\ref{fig:qpsis1} the nominal
  power-counting size for $C_{\psi S1}=-\xi m_t/16\pi^2 v^2$ is used,
  whereas $c_{ggh}$ Fig.~\ref{fig:cggh} is taken to be smaller than
  its power-counting value of ${\cal O}(1)$ due to experimental
  constraints. It is thus plausible that also $C_{\psi S1}$ is realistically
  smaller than show in Fig.~\ref{fig:qpsis1}, in line with its
  role as a subleading effect. 
\item The correction from $C_{GU1}$ (Fig.~\ref{fig:qgu1}) is enhanced
  by a factor $s^2$ with respect to the SM, at the amplitude level.
  As we discussed, this behaviour should
  not be taken at face value if the EFT is to remain applicable.
  In fact, the chiral-dimension six coefficient $C_{GU1}$ comes with a strong
  suppression displayed in (\ref{cpsis1cgu1}).
  For typical values of the coefficient
  and for energies below the EFT cutoff $M\sim 8\, {\rm TeV}$, the effect
  of $C_{GU1}$ on the cross-section is characteristic for a NLO correction.
  The behaviour of the correction from $C_{GU2}$ is expected to be similar.
\item In Fig.~\ref{fig:goldstonezl} we compare the SM amplitudes for
  $gg\to \varphi^0 \varphi^0$ and $gg\to Z_L Z_L$ using the formulae
  in~\cite{Glover:1988rg}. We find excellent agreement between the two,
  the deviation for $\sqrt{s}=500 \, \text{GeV}$ is already below 1\%.
  This is to be expected since the deviations from the Goldstone limit
  scale as $\sim m_Z^2/s$ and thus become negligible for large $\sqrt{s}$.
\end{itemize}

\section{Conclusions}
\label{sec:concl}

We have carried out a systematic analysis of EFT corrections to
longitudinal $Z$-boson pair production via gluon fusion, emphasizing the
kinematic regime in which the Higgs boson is highly off-shell.
The most appropriate EFT for this process is the electroweak chiral
Lagrangian (nonlinear EFT), whose power counting is organized as a loop
expansion, captured by counting chiral dimensions.
The process $gg\to Z_LZ_L$ is generated at the one-loop level.
We show that the EFT effects at this (leading) order depend on
three anomalous couplings, which combine to two independent parameters.
We then identify the
NLO operators contributing at two-loop order in the chiral counting and
outline the additional terms that a complete NLO calculation would require.
In the Goldstone limit, the leading‐order amplitude admits a compact
parametrization in terms of two form factors. We present explicit
expressions for these form factors and derive subleading corrections
induced by local NLO operators.
To validate our power‐counting assumptions, we study several illustrative
new‐physics scenarios that generate specific anomalous couplings.
Although some NLO contributions exhibit a pronounced growth with the
partonic center‐of‐mass energy $s$, we demonstrate that they remain
subdominant throughout the domain of validity of the EFT. Furthermore,
we find that the local anomalous Higgs–gluon coupling $c_{ggh}$, which
enters at leading order, provides an excellent approximation for
heavy‐resonance–mediated new physics. A brief phenomenological study
confirms that $c_{ggh}$ dominates the new‐physics effects at large
center-of-mass energies.
Beyond the specific case of $gg\to Z_LZ_L$, the present analysis
exemplifies how the power counting of the electroweak chiral Lagrangian
implies a systematic treatment of anomalous couplings
in general Higgs-boson related processes.

\newpage

\section*{Acknowledgements}

This work is supported in part by the Deutsche
Forschungsgemeinschaft (DFG, German Research Foundation)
under Germany’s Excellence Strategy – EXC-2094 – 390783311.

\appendix

\numberwithin{equation}{section}

\section{\boldmath Loop functions}
\label{sec:loopf}


The scalar integrals needed for the calculation are the three-point function
and the four-point function.
The general three point function is given by 
\begin{align}
  C(l_1^2,l_2^2,(l_1+l_2)^2;m_1^2,m_2^2,m_3^2) =
  \int \frac{d^4k}{i\pi^2} \frac{1}{\left[k^2-m_1^2\right]
    \left[(k+l_1)^2-m_2^2\right]\left[(k+l_1+l_2)^2-m_3^2\right]}
\end{align}
We need the special case
\begin{align}
 C(q^2) = C(0,0,q^2;m^2,m^2,m^2)=\frac{1}{2q^2} f\left( \tau\right)
\end{align}
where $\tau = q^2 / 4m^2$ and
\begin{align}
 f(\tau) = \begin{cases}
   \ln^2  \frac{\sqrt{1-\tau^{-1}}+1}{\sqrt{1-\tau^{-1}}-1} &
                       \text{for $\tau<0$}\\
   -4 \arcsin^2 \sqrt{\tau} & \text{for $0 \leq \tau \leq 1$} \\
   \left[ \ln \frac{1+\sqrt{1-\tau^{-1}}}{1-\sqrt{1-\tau^{-1}}} -i\pi\right]^2 &
       \text{for $\tau>1$}
\end{cases}  
\end{align}

The general scalar four-point function is given by
\begin{align}
  &D(l_1^2,l_2^2,l_3^2,l_4^2,(l_1+l_2)^2,(l_2+l_3)^2;m_1^2,m_2^2,m_3^2,m_4^2)=
  \nonumber
  \\ &= \int \frac{d^4 k}{i \pi^2} \frac{1}{\left[k^2-m_1^2\right]
    \left[(k+l_1)^2-m_2^2\right]\left[(k+l_1+l_2)^2-m_3^2\right]
    \left[(k+l_1+l_2+l_3)^2-m_4^2\right]}
\end{align}
Again we need a special case, 
\begin{align}
	D(q^2,r^2) = D(0,0,0,0,q^2,r^2;m^2,m^2,m^2,m^2) = \frac{2}{q^2 r^2} \frac{1}{\beta_2(\tau,\sigma)} g(\tau,\sigma)
\end{align}
where we defined 
\begin{align}
	\tau = \frac{q^2}{4m^2}, \quad  \sigma = \frac{r^2}{4m^2}, \quad \beta_1(z) = \sqrt{1-z^{-1}}, \quad \beta_2(y,z) = \sqrt{1-y^{-1}-z^{-1}}
\end{align}
and
\begin{align}
	g(\tau,\sigma) = I_{1}(\tau,\sigma) + I_{1}(\sigma,\tau) +I_{2}(\tau,\sigma) - \frac{\pi^2}{2}
\end{align}
with 
\begin{align}
	I_{1}(\tau,\sigma) =& \; 2 \text{Li}_2\left(\tau \left[\beta_1(\sigma)-1\right]\left[\beta_1(\sigma)-\beta_2(\tau,\sigma)\right]\right) -2 \text{Li}_2\left(-\tau \left[\beta_1(\tau)-1\right]\left[\beta_1(\tau)-\beta_2(\tau,\sigma)\right]\right) \nonumber \\
	&-\ln^2\left(\tau \left[\beta_1(\sigma)+1\right]\left[\beta_1(\sigma)-\beta_2(\tau,\sigma)\right]\right)
\end{align}
and
\begin{align}
	I_{2}(\tau,\sigma) = I_{2}(\sigma,\tau)= & \ln\left(- \sigma \left[\beta_1(\tau)-\beta_2(\tau,\sigma)\right]^2\right)\ln\left(- \tau \left[\beta_1(\sigma)-\beta_2(\tau,\sigma)\right]^2\right) \nonumber \\
	&+2 \ln^2 \left(\tau \left[\beta_1(\tau)+\beta_2(\tau,\sigma)\right]\left[\beta_1(\sigma)-\beta_2(\tau,\sigma)\right]\right)
\end{align}
The above expression for $g(\tau,\sigma)$ is immediately applicable in the
region $\tau,\sigma <0 $. For $\tau,\sigma >0$ it holds with the prescription
$\tau \to \tau + i\eta$ and $\sigma \to \sigma + i\eta$ respectively.


\end{document}